\documentclass[10pt, conference, compsocconf]{IEEEtran}

\ifCLASSINFOpdf
\else
\fi

\usepackage{url}
\usepackage{color}
\usepackage{multirow}
\usepackage{graphicx}
\usepackage{subfigure,tabularx}
\usepackage{amsfonts,amsmath,color,amssymb,amsxtra,graphicx,floatflt}
\usepackage{hhline}
\usepackage{booktabs,multirow}
\usepackage{fancyhdr}

\usepackage{rotating}

\newlength{\bibitemsep}
\newlength{\bibparskip}
\let\oldthebibliography\thebibliography
\renewcommand\thebibliography[1]{%
  \oldthebibliography{#1}%
  \setlength{\parskip}{\bibitemsep}%
  \setlength{\itemsep}{\bibparskip}%
}

\fancypagestyle{IEEE_Green_open_access_footer}{
  \fancyhf{}
  \fancyhead[C]{\footnotesize \copyright2019 IEEE. Personal use of this material is permitted. Permission from IEEE must be obtained for all other uses, in any current or future media, including reprinting/republishing this material for advertising or promotional purposes, creating new collective works, for resale or redistribution to servers or lists, or reuse of any copyrighted component of this work in other works. DOI: 10.1109/GIOTS.2019.8766431}

}

\begin{document}

\title{Atomic Services:\\sustainable ecosystem of smart city services through pan-European collaboration}

\author{\IEEEauthorblockN{Flavio Cirillo*, Detlef Straeten}
\IEEEauthorblockA{NEC Laboratories Europe\\
Heidelberg, Germany\\
\{flavio.cirillo,detlef.straeten\}@neclab.eu\\
*also: Dept. of Electrical Eng. and Information Tech.\\
University of Naples Federico II, \\
Naples, Italy}
\\
\\
\IEEEauthorblockN{Luis Diez, Ignacio Elicegui Maestro}
\IEEEauthorblockA{Communications Engineering Department\\
University of Cantabria\\
Santander, Spain\\
\{ldiez,iemaestro\}@tlmat.unican.es}

\and

\IEEEauthorblockN{David G\'omez, Jose Gato}
\IEEEauthorblockA{Atos Research and Innovation\\
Atos Spain\\
Madrid, Spain\\
\{david.gomez,jose.gato\}@atos.net\\
\hspace{4cm}\\
\hspace{4cm}}
\\
\\
\IEEEauthorblockN{Reza Akhavan}
\IEEEauthorblockA{Future Cities Catapult\\
London, UK\\
rakhavan@futurecities.catapult.org.uk}
\and
%\IEEEauthorblockN{Luis Diez, Ignacio Elicegui Maestro}
%\IEEEauthorblockA{Communications Engineering Department\\
%University of Cantabria\\
%Santander, Spain\\
%\{ldiez,iemaestro\}@tlmat.unican.es\\
%\hspace{4cm}\\
%\hspace{4cm}}
%\and
%\IEEEauthorblockN{Reza Akhavan}
%\IEEEauthorblockA{Future Cities Catapult\\
%London, UK\\
%rakhavan@futurecities.catapult.org.uk\\
%\hspace{8cm}\\
%\hspace{4cm}}
}

\IEEEoverridecommandlockouts
\IEEEpubid{\makebox[\columnwidth]{978-1-7281-2171-0/19/\$31.00 \copyright~2019~IEEE }
\hspace{\columnsep}\makebox[\columnwidth]{ }}

\maketitle
\thispagestyle{IEEE_Green_open_access_footer}

\begin{abstract}
In a world with an ever increasing urbanization, governance is investigating innovative solutions to sustain the society evolution. Internet-of-Things promises huge benefits for cities and the proliferation of smart city deployments demonstrates the common acceptance of IoT as basis for many solutions.
The city pilots developments occurred in parallel and with different designs thus creating fragmentation of IoT. The European project SynchroniCity aims to synchronize 8 smart cities to establish a shared environment fostering a self-sustained business growth.
In this article we present the collaborative methodology and shared efforts spent towards the creation of a common ecosystem for the development of smart city services. Our design evolves around the concept of "atomic services" that implements a single functional block to be composed for full-fledged smart city services. This creates opportunities for diverse stakeholders to participate to a global smart cities market.
The methodology and outcome of our efforts will be followed by 10 new cities globally, thus expanding the market range for IoT stakeholders.
\end{abstract}


\IEEEpeerreviewmaketitle

%%%%%%%%%%%%%%%%%%%%%%%%%%%%%%%%%%%%%%%%%
%%%%% Introduction %%%%%%%%%%%%%%%%%%%%%%
%%%%%%%%%%%%%%%%%%%%%%%%%%%%%%%%%%%%%%%%%

\section{Introduction}

The Internet-of-Things (IoT) concept has been widely adopted in the context of smart cities due to the huge potential benefits for the ever increasing urbanized society, in terms of services for the citizens and for business growth. The proliferation of smart cities' infrastructure is extensive around the world~\cite{SILVA2018697}. The developments of smart cities happened in parallel following different approaches and designs. Consequently this led to the isolation of IoT segments between cities, and even among domains within the same city~\cite{vital2015}. To overcome this situation the SynchroniCity project\footnote{\url{https://synchronicity-iot.eu/}} has the purpose of ``synchronizing'' IoT infrastructures of smart cities in order to address the ``city lock-in'' barrier. Eight cities in Europe (i.e. Milan, Santander, Porto, Helsinki, Eindhoven, Antwerp, Manchester, and Carouge) have been chosen as pilot cities (named reference zones - RZs) for being harmonized~\cite{cirillo2019}. RZs are jointly developing and instantiating an overlay infrastructure on top of existing systems. The outcome of this project will be then leveraged soon by 10 new cities joining SynchroniCity and afterwards within the Open \& Agile Smart Cities (OASC)\footnote{https://oascities.org/} association by the more than 100 cities worldwide involved.

On top of the synchronized IoT infrastructures, several IoT applications are conceived for the betterment of citizens life. The applications are to be developed in cooperation amongst city pilots and other stakeholders (such as academia and SMEs), breaching the barrier of monolithic vertical developments only sustainable for big companies~\cite{AIOTI-WG02}. The idea is to build smart city applications by composing small services~\cite{SynCity-D3-2}, namely atomic services\footnote{online available:\\ https://gitlab.com/synchronicity-iot?filter=atomic+service}, where each constitutes a well-defined functional block. Internet-of-Things ecosystem, indeed, is envisioned to be not only about data but also services~\cite{biotope}. This approach embraces the concepts of service oriented architecture (SOA)~\cite{abu-matar}, opening opportunities also to small stakeholders in the IoT market~\cite{westerlund2014}. 

Because of the large pilot nature of SynchroniCity project, we could move from the collaborative creation approach within the same urban area~\cite{schaffers2011} to a pan-European multi-actor collaboration. The project partners planned to implement 11 smart city applications, namely "city services", based on the exploitation of homogenized and similar IoT data sources and categorized in three application themes~\cite{SynCity-D3-1}: ``human-centric traffic management'', ``multi-modal transportation'', and ``community policy suite''.
For those applications a total of 8 atomic services have been identified by following two approaches. The bottom-up approach started from the requirements of the city services. We have compiled and distributed a questionnaire to prioritize requirements, and to discover previous experiences and available assets. The questionnaire was answered by representative of all the 8 pilot cities plus an additional experienced SME. From this phase we have identified 3 available assets (e.g. geographical routing, visualization services) that were adopted as atomic services to address cities requirements. Their integration 
was performed in an agile manner with continuous synchronization with interfaces evolution (both with the SynchroniCity infrastructure and between atomic services). This resulted with the identification of 3 new needed atomic services (e.g. for handling General Transit Feed Specification - GTFS data).
Applications have then being designed on top of these initial set of atomic services.
During the second, top-down, approach we analyzed the common points between applications architectures. Thus we have shared design and implementation efforts with definition of re-usable functional blocks, resulting with other 2 atomic services (i.e. data analytics based estimator service). 
The services were then adopted and integrated with other customized components, specific of the smart city application. We demonstrate that huge percentage (up to 87,5\%) of the IoT applications to be developed was addressed collaboratively. 
Current developments of the SynchroniCity projects, with 10 new cities and 16 new city services, will certainly enrich the roster with additional atomic services.

The design of the adopted collaborative methodology is introduced in Section~\ref{sec:atomicservices}. The analysis of the gathered information is described in Section~\ref{sec:reqanalysis}. Atomic services embodiments, and an actual combination and customization for the implementation of one of the smart city applications, is presented in Section~\ref{sec:embodiments}. Finally 
Section~\ref{sec:results} shows the evaluation of the effectiveness of the followed approach.

%%%%%%%%%%%%%%%%%%%%%%%%%%%%%%%%%%%%%%%%%
%%%%% CoCreation %%%%%%%%%%%%%%%%%%%%%%
%%%%%%%%%%%%%%%%%%%%%%%%%%%%%%%%%%%%%%%%%
\section{Atomic Services: collaborative approach to IoT application creation}
\label{sec:atomicservices}

Cities within SynchroniCity cooperated for the creation of ``city services'' as direct demonstration of a shared ecosystem. A city service is an application built for citizens targeting their needs with the aim of ``smartify'' the life within urban area. The eleven planned city services are categorized under three application themes~\cite{SynCity-D3-1}: ``human-centric traffic management'' improves cycling experience and safety; ``multi-modal transportation'' provides city travellers attractive alternative to private vehicle such as the combination of integrate public transport and shared mobility services; ``community policy suite'' assists policy maker with an intelligence layer for city governance.

\begin{figure}[h]
\centering
\includegraphics[width=0.9\linewidth,trim={0cm 2.5cm 11.5cm 0cm},clip]{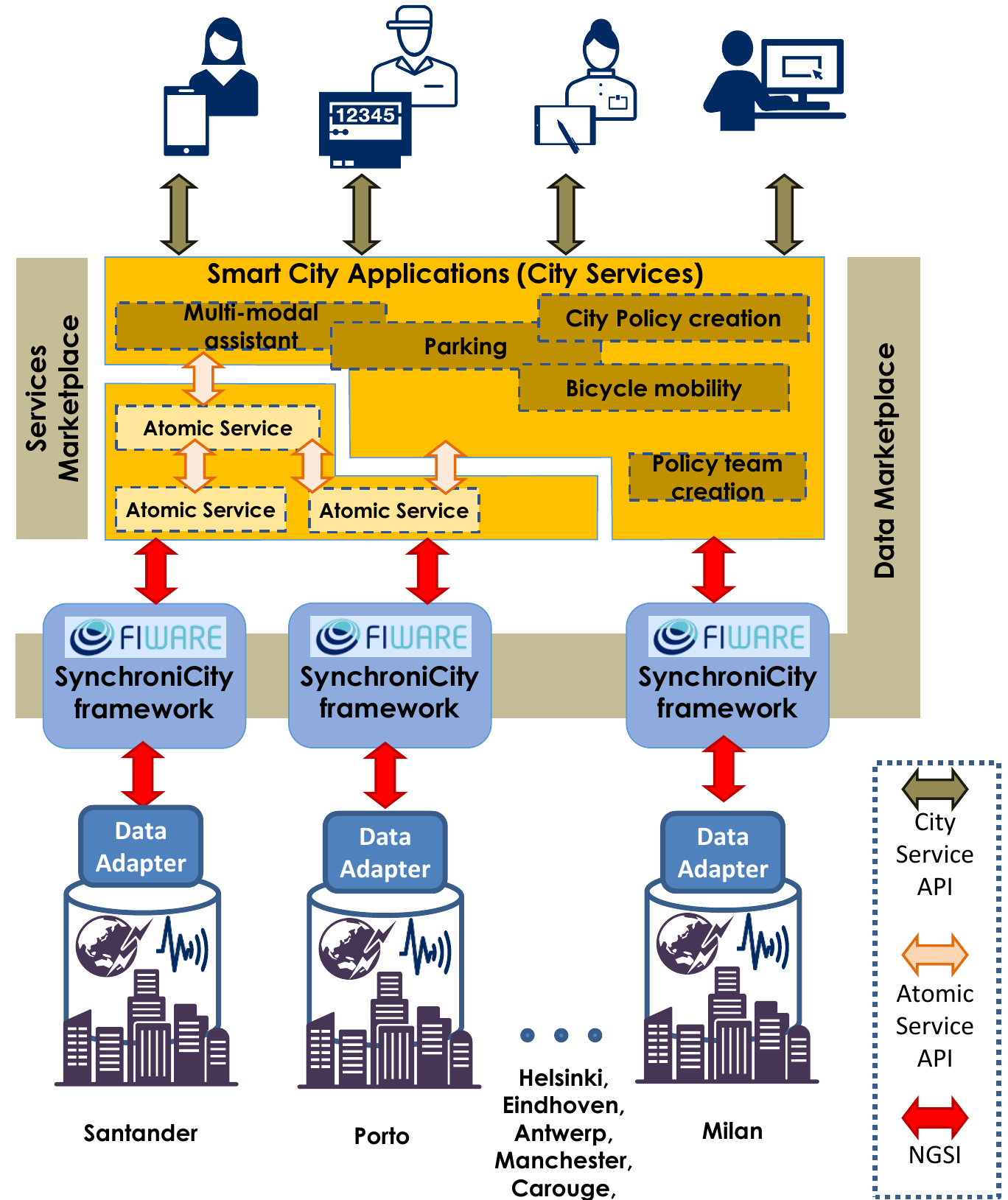}
\caption{City services and atomic services shared ecosystem.}
\label{fig:architecture}
\end{figure}

As collaborative approach we have chosen to build city services on top of components named ``atomic services\footnote{Formerly known also as ``baseline services''}''. These components are single functional blocks using data to return any kind of feature, either managing, enriching, joining or filtering the input data from the SynchroniCity infrastructure, and/or data coming from other sources. Atomic services have similarity with the concept of microservice in the fact of being a self-contained piece of software targeting a specific task. Nevertheless, an important characteristic of atomic service is that it must be re-usable (hence generic) by more than one city services. Therefore, whilst an atomic service can be used as a microservice of a certain city service, the contrary cannot be always true. For example, as a database service is not conceived primarily as a microservice but only used in different manners as such in microservice design~\cite{Messina2016-DBaaMS}, so an atomic service is not a microservice per se. 
Atomic services are offering facilities, on the one hand, to ease city services development and reduce the time-to-market and, on the other hand, to share efforts among city services teams. The nature of the atomic services facilitates reference zone customization and stimulates replication based on common solutions and designs. These common solutions prevent both “vendor lock-in” from the cities’ perspective and “city lock-in” from the perspective of city service and middleware providers. Figure~\ref{fig:architecture} depicts the overall concept of the shared ecosystem. Operational smart city systems are homogenized with the FIWARE-based SynchroniCity framework and data are then exposed with same interfaces and data-models (red arrows). City services might use none, one or a combination of the features offered by atomic services. Data and services marketplaces expose valuable assets to third parties enabling an attractive and sustainable ecosystem.

The design of the city services started with the identification of stakeholders and functional requirements (FRs) for each of the application themes~\cite{SynCity-D3-1}. As methodology to define atomic services from FRs we grouped them in small subsets, and for each of the subset we have assigned an atomic service to address the related FRs.
These components are cooperatively implemented among cities (involving also their technical partners), and the final atomic services shared with the others. 

In total, at the time of writing this work (Feb. 2019), there have been identified up to 8 atomic services, categorized as follow: 2 data prediction services for parking, and traffic flow; 2 visualization services, that are a dashboard and a metrics visualizer; 1 data elaboration service for geographic route calculation; and 3 data transformation services related to generate or retrieve General Transit Feed Specification (GTFS) and GTFS-RT from NGSI. Due to the active project and long run ambition of the SynchroniCity project, it is foreseen to have more atomic services in the future.

\subsection{Questionnaire} 
The identified FRs cover all the aspects of an application theme, and typically not all are required by a specific city service. Moreover cities and project partners have previous experiences with smart cities applications, thus already existing components might be re-used simply off-the-shelf, or with minimal integration effort.

For these motivations we have created a questionnaire (see Table~\ref{tab:questionnaire}) with the following targets: 1) identify previous experiences with the application themes, 2) identify re-usable software components, 3) identify possibility to interact with the SynchroniCity infrastructure (read-only or read-write), 4) identify which packaging tool is preferred for sharing components, 5) identify license foreseen for the atomic/city services to be implemented, 6) prioritize application themes FRs for each city.

\begin{table}[htbp]
\caption{Atomic services and city services questionnaire}
\label{tab:questionnaire}
\vspace{-2em}
\begin{center}
\setlength\tabcolsep{1.5pt}
\begin{tabular}{|c|c|c|c|c|}

\hline

\multicolumn{4}{|c|}{\textbf{Question}} & 
\multicolumn{1}{p{0.1\linewidth}|}{\textbf{Answer}} \\

\hline
\hline

\multicolumn{4}{|l|}{\begin{tabular}{@{}l@{}} $ \bullet $ Do you have already implemented an application satisfying\\\hspace{1em}the proposed use-cases~\cite{SynCity-D3-1}? \end{tabular}} & 
\multicolumn{1}{p{0.1\linewidth}|}{} \\

\multicolumn{4}{|l|}{\begin{tabular}{@{}l@{}} $ \bullet $ If yes at the above question, can you please  provide an\\\hspace{1em}architecture overview (sub-components, atomic services)? \end{tabular}} & 
\multicolumn{1}{p{0.1\linewidth}|}{} \\

\hline

\multicolumn{4}{|l|}{\begin{tabular}{@{}l@{}} $ \bullet $ Do you have any related atomic services already implemented? \end{tabular}} & 
\multicolumn{1}{p{0.1\linewidth}|}{} \\

\multicolumn{4}{|l|}{\begin{tabular}{@{}l@{}} $ \bullet $ If yes at the above question, are you willing to share them\\\hspace{1em}(OpenSource, Freeware, Licensing etc.)? \end{tabular}} & 
\multicolumn{1}{p{0.1\linewidth}|}{} \\

\multicolumn{4}{|l|}{\begin{tabular}{@{}l@{}} $ \bullet $ If yes at the above question, can you please  provide an\\\hspace{1em}architecture overview for each atomic services you possess? \end{tabular}} & 
\multicolumn{1}{p{0.1\linewidth}|}{} \\

\hline

\multicolumn{4}{|l|}{\begin{tabular}{@{}l@{}} $ \bullet $ Do you envision any additional atomic service to share across\\\hspace{1em}application of the same theme and/or across application of\\\hspace{1em}different themes? Please give a brief description/suggestion. \end{tabular}} & 
\multicolumn{1}{p{0.1\linewidth}|}{} \\

\hline

\multicolumn{4}{|l|}{\begin{tabular}{@{}l@{}} $ \bullet $ Is the NGSI data from the IoT infrastructure read-only towards\\\hspace{1em}the atomic services or the atomic services will be capable\\\hspace{1em}of writing new data? (e.g. new entities or commands like in\\\hspace{1em}the case of traffic light controlling) \end{tabular}} & 
\multicolumn{1}{p{0.1\linewidth}|}{} \\

\hline

\multicolumn{4}{|l|}{\begin{tabular}{@{}l@{}} $ \bullet $ Which methods and related tools are desired for atomic service\\\hspace{1em}deployment (e.g. docker, debian package)? \end{tabular}} & 
\multicolumn{1}{p{0.1\linewidth}|}{} \\

\hline

\multicolumn{4}{|l|}{\begin{tabular}{@{}l@{}} $ \bullet $ Will the new code your are going to implement for\\\hspace{1em}SynchroniCity be reusable? (Open Source, Freeware etc.)?\end{tabular}} & 
\multicolumn{1}{p{0.1\linewidth}|}{} \\

\hline

\multirow{4}{*}{\begin{tabular}{@{}l@{}}
$ \bullet $ Which FR your application\\
\hspace{1em}shall satisfy? (Priority 0 to 5:
\\\hspace{1em}0 for FR not needed, 5 for
\\\hspace{1em}FR of utmost importance)
\\\hspace{1em}In case you have atomic
\\\hspace{1em}services implemented, which
\\\hspace{1em}FR does it fully or partially 
\\\hspace{1em}satisfy/support?\end{tabular}} & 
FR-1 & 
\begin{tabular}{@{}c@{}}Priority\\(0-5) \end{tabular} & 
\begin{tabular}{@{}c@{}}Already satisfied;\\Partially satisfied;\\not implemented;\\n.a.\end{tabular} &  \\
\cline{2-5}
& FR-2 & \begin{tabular}{@{}c@{}}Priority\\(0-5) \end{tabular} & \begin{tabular}{@{}c@{}}Satisfaction\\level\end{tabular} & \\
\cline{2-5}
& ... & \begin{tabular}{@{}c@{}}Priority\\(0-5) \end{tabular} & \begin{tabular}{@{}c@{}}Satisfaction\\level\end{tabular} & \\
\cline{2-5}
& FR-n & \begin{tabular}{@{}c@{}}Priority\\(0-5) \end{tabular} & \begin{tabular}{@{}c@{}}Satisfaction\\level\end{tabular} & \\
\hline

\end{tabular}
\end{center}
\end{table}

\subsection{Functional Requirements (FRs) analysis.} 

We have distributed the questionnaire to the 8 cities plus additional partners that have worked with other cities (some of the questions were not applicable to the latter group). The questionnaire was answered by all the involved cities for each application themes, with additional answers by other cities and partners of the project. Table~\ref{tab:selectedfrs} shows the cities that answered the questionnaires; between brackets are the additional cities not directly involved in the related application themes. In total 9 partners have participated, categorized in 5 municipalities, 2 SMEs, and 2 academia/research foundations.
From the answers we have filtered and prioritized the functional requirements with the following criteria: 1) threshold to filter out FRs not to be addressed because not of common interest: more than 2 cities with priority equal or higher than 3; 2) filter out FRs already supported by the data producers and IoT framework; 3) rank by priorities average. 

\begin{table}[htbp]
\caption{Functional Requirements (FRs) of common interest}
\label{tab:selectedfrs}
\vspace{-1em}
\begin{center}
\setlength\tabcolsep{1.5pt}
\begin{tabular}{|l|c|c|c|c|}

\hline
\textbf{Application Theme}
& \textbf{\begin{tabular}{@{}c@{}}Cities answering\\the quest.\end{tabular}} 
& \textbf{\begin{tabular}{@{}c@{}}Total\\FRs\end{tabular}} 
& \textbf{\begin{tabular}{@{}c@{}}Selected\\FRs\end{tabular}}
& \textbf{\begin{tabular}{@{}c@{}}FRs to\\address\end{tabular}}\\
\hline
\hline
Human-centric traffic mgmt. & 
\begin{tabular}{@{}c@{}}Ant, Ein, Mil;\\(Car, San)\end{tabular}
& 12 & 8 & 7 \\
\hline
Multi-modal transportation & 
\begin{tabular}{@{}c@{}}Hel, Mil, Por,\\San; (Car)\end{tabular}
& 33 & 28 & 19 \\
\hline
Community Policy Suite & 
\begin{tabular}{@{}c@{}}Car, Man, Por;\\(San)\end{tabular}
& 9 & 8 & 8 \\
\hline

\end{tabular}
\end{center}

\end{table}

In particular, for point 2), we have evaluated the SynchroniCity framework~\cite{SynCity-D2-10} and the legacy IoT systems involved. For instance, the security layer of the IoT framework fully addresses three FRs (plus partially a fourth one). In fact, the Single Sign-On (SSO) approach is envisioned for all the city services that are developed following the SynchroniCity philosophy of a shared ecosystem. This, on the one hand, relieves the application developer from the burden of managing credentials, and on the other hand, the city service users (e.g. citizens) to create a new user for every new city service. Another FR was addressed by the geographical data query and subscription methods offered by the infrastructure. Finally the cities IoT providers together with the adoption of a common standard, such as OMA NGSI~\cite{oma_ngsi}, and common data models, such as FIWARE data models\footnote{https://www.fiware.org/developers/data-models/}, were already addressing six data related requirements. Table~\ref{tab:selectedfrs} summarizes the filtered FRs, a more detailed list of selected FRs can be viewed in~\cite{SynCity-D3-2}.

%%%%%%%%%%%%%%%%%%%%%%%%%%%%%%%%%%%%%%%%%
%%%%% RequirementsAnalysis %%%%%%%%%%%%%%%%%%%%%%
%%%%%%%%%%%%%%%%%%%%%%%%%%%%%%%%%%%%%%%%%
\section{Requirements Analysis}
\label{sec:reqanalysis}

The identification of atomic services followed two different approaches: bottom-up and top-down. The bottom-up approach started from making an inventory of available software assets that address at least one FR. Afterwards, we chose the ones best suited for the IoT environment on which those assets are going to work, or with better offered features. The bottom-up approach encompasses also the analysis of the integration with the IoT infrastructure (e.g. data format mediation). The integration might be realized either with customization of the available assets or with the identification of new services.
The second approach (top-down) started from the comparison between the designs of city services. Functional blocks that appear in multiple designs are selected to become atomic services and to be implemented with shared efforts.

\begin{table}[htbp]
\caption{Atomic Services selected with different approaches.}
\label{tab:atomicservices}
\vspace{-1em}
\begin{center}
\setlength\tabcolsep{1pt}
\begin{tabular}{|c|p{22pt}|p{22pt}|p{22pt}|p{22pt}|p{22pt}|p{22pt}|>{\centering}p{22pt} |p{22pt}|}

\hline
\textbf{Category} &  
\multicolumn{2}{c|}{\begin{tabular}{@{}c@{}}data\\prediction\end{tabular}} & 
\multicolumn{1}{c|}{\begin{tabular}{@{}c@{}}data\\eval.\end{tabular}} &
\multicolumn{2}{c|}{visual.} &
\multicolumn{3}{c|}{\begin{tabular}{@{}c@{}}data\\tanformation\end{tabular}} \\
\hline
\begin{tabular}{@{}c@{}}\textbf{Atomic}\\\textbf{Service}\end{tabular} & 
\hspace{0.2em}\rotatebox{90}{\begin{tabular}{@{}c@{}}parking\\estimator\end{tabular}} & 
\hspace{0.2em}\rotatebox{90}{\begin{tabular}{@{}c@{}}traffic flow\\estimator\end{tabular}} & 
\hspace{0.8em}\rotatebox{90}{\begin{tabular}{@{}c@{}}routing\end{tabular}} & 
\hspace{0.8em}\rotatebox{90}{\begin{tabular}{@{}c@{}}dashboard\end{tabular}} & 
\hspace{0.2em}\rotatebox{90}{\begin{tabular}{@{}c@{}}metrics\\visualizer\end{tabular}} & 
\hspace{0.2em}\rotatebox{90}{\begin{tabular}{@{}c@{}}gtfs\\fetcher\end{tabular}} & 
\hspace{0.2em}\rotatebox{90}{\begin{tabular}{@{}c@{}}ngsi2gtfs\end{tabular}} & 
\hspace{0.2em}\rotatebox{90}{\begin{tabular}{@{}c@{}}gtfs-rt\\from ngsi\end{tabular}} \\
\hline
\begin{tabular}{@{}c@{}}\textbf{Bottom-up}\\\textbf{Approach}\end{tabular} & 
 & & \hspace{1em}\checkmark & \hspace{1em}\checkmark & \hspace{1em}\checkmark & \hspace{1em}\checkmark & \checkmark  & \hspace{1em}\checkmark  \\
\hline
\begin{tabular}{@{}c@{}}\textbf{Top-Down}\\\textbf{Approach}\end{tabular} &  
\hspace{1em}\checkmark & \hspace{1em}\checkmark & & & & & & \\
\hline

\end{tabular}
\end{center}
\end{table}

\subsection{Bottom-up approach}

The questionnaire answers gave an overall view about available assets, that we could exploit to start designing the applications. For example, we have identified two candidates as routing service to be used in two of the three application themes (i.e. human-centric traffic management and multi-modal transportation): OpenTripPlanner (OTP)\footnote{http://www.opentripplanner.org/} and BRouter\footnote{http://brouter.de/brouter/}. The choice fell upon OTP since it covers more FRs compared with BRouter, in particular regarding the possibility of including real-time information (such as road accidents and air quality) in the routing calculation. Furthermore OTP was already successfully tested on a previous smart city application of a partner city (i.e. Helsinki\footnote{https://digitransit.fi/}). The chosen routing service was not ready to be used with the SynchroniCity infrastructure, since it is accepting as input data packaged in General Transit Feed Specification (GTFS)\footnote{\url{https://gtfs.org/}} and GTFS-Realtime (GTFS-RT) whereas the IoT systems are all following the NGSI standards and FIWARE data models. Being FIWARE an open community, this incompatibility of format has been resolved with the definition of new data models\footnote{FIWARE Urban Mobility data model: https://fiware-datamodels.readthedocs.io/en/latest/UrbanMobility/doc/introduction/} specifically devoted to map and/or carry GTFS data within NGSI. For that reason, three new atomic services have been designed: the GTFS Fetcher discovers and extracts GTFS static data from the IoT platform; the NGSI Urban Mobility to GTFS generates GTFS files from the discovered NGSI urban mobility data laying in the IoT system; the GTFS-RT from NGSI, similarly, generated GTFS-RT files.   Following the philosophy of re-usability and IoT service ecosystem creation, we have designed these atomic services as standalone, instead of glue code wrapping the OTP. Thus these GTFS related services can be adopted in other context requiring NGSI and GTFS.

Two other tools have been identified related to data visualization. The first one, namely SmartCities Dashboard, displays IoT devices and data on a map. The second tool, i.e. the Metrics Visualizer (based on Grafana\footnote{https://grafana.com/}), visualizes timeseries of IoT data. Both of them were already integrated and tested with the SynchroniCity infrastructure.

\subsection{Top-down approach}

With the identified assets some aspects of the actual city services were not addressed. For that reason we proceeded with the design of the final applications. In particular there were missing data inputs regarding real-time parking situations aggregated by city neighbourhoods, parking situations forecasts, and traffic flow status within urban area. For those purposes we envisaged two atomic services performing data analytics on time-series data: Parking Probability Estimator, and Traffic Flow Estimator.

Section~\ref{sec:embodiments} depicts the two estimator atomic services. In addition, it presents a concrete example of atomic services composition for the realization of the city service that will be operational in the city of Santander (Spain) 

%%%%%%%%%%%%%%%%%%%%%%%%%%%%%%%%%%%%%%%%%
%%%%% Examples %%%%%%%%%%%%%%%%%%%%%%
%%%%%%%%%%%%%%%%%%%%%%%%%%%%%%%%%%%%%%%%%
\section{Atomic Service Embodiments}
\label{sec:embodiments}

\subsection{Data analytics atomic services}

Amongst the roster of atomic services, we focus in this section on two of them, which make a thorough use of the off-the-shelf assets exposed by the different reference zones. Namely, we concentrate on parking and traffic (intensity) data to build two different so-called atomic services: one estimates the availability of a free parking spot, the other ``predicts'' the state of the traffic. On top of them, cities build their own application, which might embrace them in, for instance, a single-stop-shop.

\begin{figure}[tp]
    \centering
    \includegraphics[width=\columnwidth]{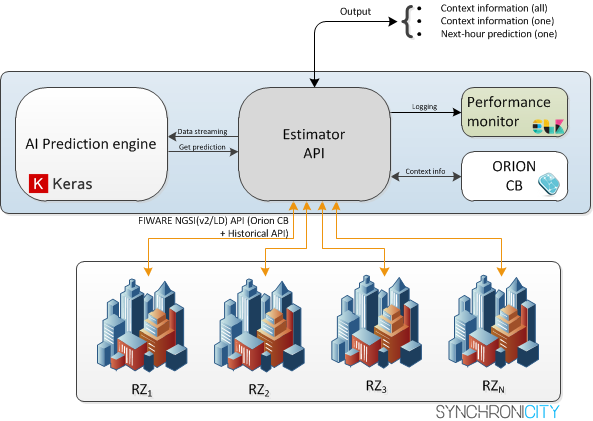}
    \caption{Parking and traffic flow estimators shared architecture}
    \label{fig:estimator}
\end{figure}

By means of an Artificial Intelligence (AI) engine, based on the Keras framework\footnote{Keras. The Python Deep Learning Library - \url{https://keras.io/}}, these two services bring the potential to go beyond the current status of the cities, alike most of the current off-the-shelf applications and services, and offer an estimation (i.e. one hour ahead in time) of what is about to happen in a particular (and queried) area. The way to address this is shown in Figure~\ref{fig:estimator}, where we can distinguish between four different components:

\begin{enumerate}
    \item \textbf{Estimator API}. This component can be seen as the communication broker. Its operation can be split into the following tasks, as reflected in the figure: 1) Harvest data from the RZs' FIWARE framework (query/response, subscription-based and historical queries); 2) send data to the AI prediction modules to train the models, and receive the output of the predictions; 3) persist and synchronize all the context information within the atomic service using Orion Context Broker\footnote{https://github.com/telefonicaid/fiware-orion}; 4) log every event happened in any of the sub-components; 5) expose (via a standalone RESTful API) the results of the predictions (and other context information).
    \item \textbf{AI Prediction Engine}. Core of the estimation operations. It deals with the typical workflow of most AI applications: data processing, learning/training and predicting/estimating.
    \item \textbf{Performance monitoring}. In order to cope with all the ``logs'' and events gathered at the Estimator API module, we rely on a Elastic Search-Kibana-Logstash (ELK) stack\footnote{Elastic - \url{https://www.elastic.co/es/elk-stack}} to keep track and analyze/quantify the performance as a whole. 
    \item \textbf{Orion CB}. 
    We rely on this component for a twofold role: on the one hand, to work as a regular context manager that aggregates the data from the RZs, on the other hand, to harvest the output of the estimator and behave as a cache-like system. 

\end{enumerate}

After all the process, end users can retrieve a new level of knowledge built upon the raw data directly gathered from the underlying RZs.

\subsection{Atomic services as building blocks for city services}

As we have seen in previous sections, atomic services are intended to provide a simple and well defined functionality.
We can build complex final services, on top of the SynchroniCity platform, as a composition of atomic services. This section describes the design of a urban routing service deployed on top of a SynchroniCity instance. 

From an overarching perspective, the whole service presents a three-layered design, as shown in Figure~\ref{fig:sc}. At the bottom, through its Context Broker, the SynchroniCity platform provides context information that is adapted, in a middle layer, by the atomic services, so can be consumed by the routing engine on top. It is worth noting that, although the routing service uses Open Trip Planner\footnote{\url{http://www.opentripplanner.org/}} (OTP) as default engine, the design allows other implementations. Advanced routing engines can be fed with static and dynamic (real-time) transportation data using GTFS and GTFS-RT.
The design in Figure~\ref{fig:sc} embraces two atomic services within the SynchroniCity framework, GTFS fetcher and GTFS-RT from NGSI, addressing static and real-time data respectively. 

The approach followed by the atomic services is different: GTFS Fetcher acts as a proxy for existing data, while GTFS-RT from NGSI is a data translator. The reason is that the GTFS static information is usually available in cities, while real-time traffic information is typically exploited using proprietary formats. For that reason, we rely on NGSI and well defined data-models to provide a uniform representation of real-time information, and it is then exposed as GTFS-RT.

In the following we briefly describe the basic functionalities of both atomic services.

\subsubsection{GTFS fetcher}
As depicted in Figure ~\ref{fig:sc}, this service is made of three components: service orchestrator, file fetcher, and routing engine plugin. First, upon configuration the orchestrator retrieves the corresponding context information, and indicates the file fetcher where the require GTFS files are located. At the same time, the context information is processed and stored to properly carry out GTFS updates (e.g. validity period is checked). Once the GTFS file is retrieved, the routing engine plugin interacts with the OTP so that it uses the updated information when providing new routes.

\subsubsection{GTFS-RT from NGSI} 
\begin{figure}
	\includegraphics[width=\columnwidth]{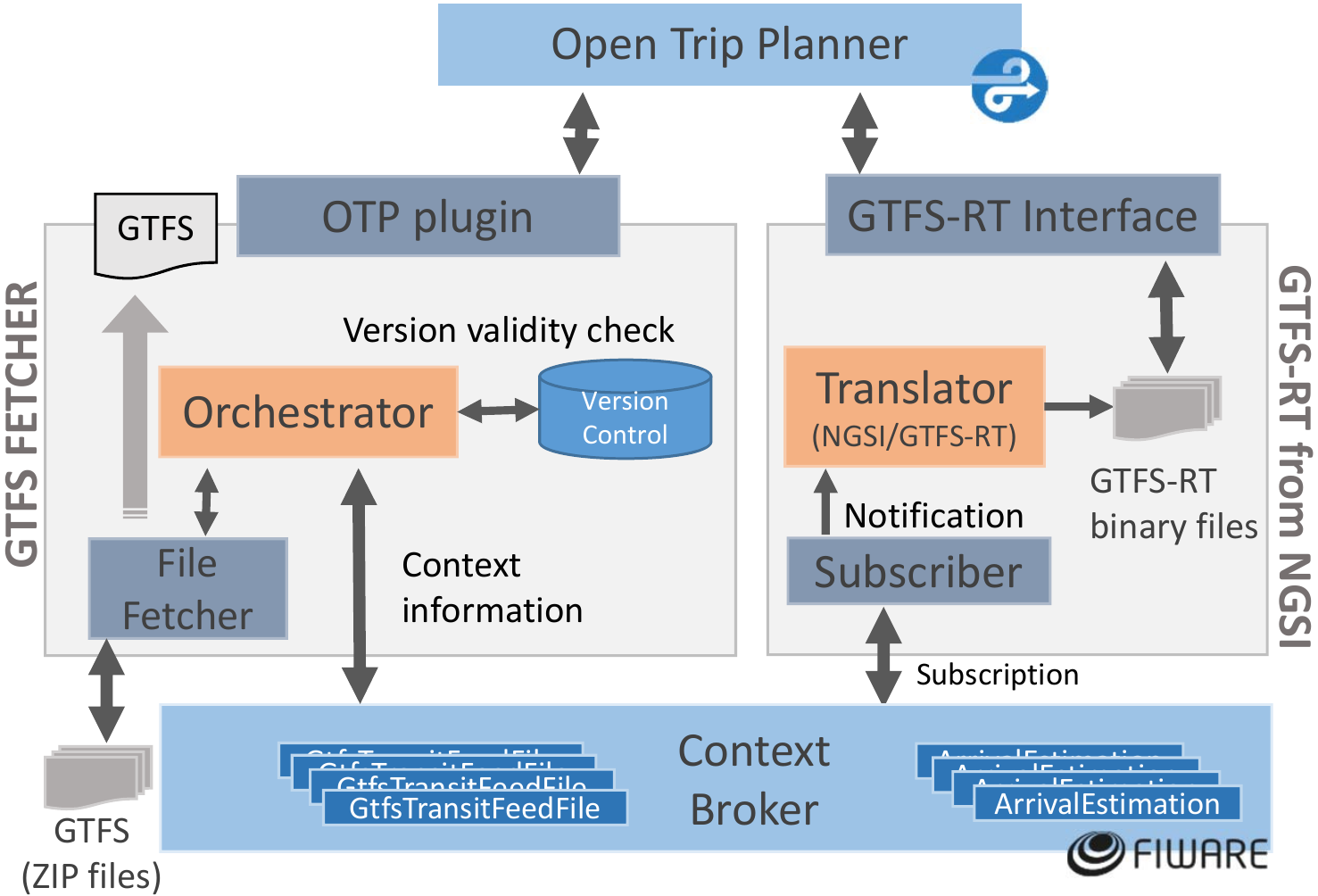}
	\caption{Routing service design\label{fig:sc}}
\end{figure}

The \emph{GTFS-RT from NGSI} atomic service creates GTFS-RT feeds from NGSI entities. As shown in Figure~\ref{fig:sc} a subscriber module creates subscriptions to the context broker to get notified when the corresponding entities are updated. Then, the notification is propagated to a translator component, that is the central part of the service. This component creates or updates GTFS-RT binary file with the new notified information, using protocol buffer\footnote{\url{https://github.com/protocolbuffers/protobuf}}. Finally, the generated GTFS-RT feeds can be consumed through a REST interface, making the final service totally agnostic of the SynchroniCity integration.

%%%%%%%%%%%%%%%%%%%%%%%%%%%%%%%%%%%%%%%%%
%%%%% Result %%%%%%%%%%%%%%%%%%%%%%
%%%%%%%%%%%%%%%%%%%%%%%%%%%%%%%%%%%%%%%%%
\section{Evaluation}
\label{sec:results}

\begin{figure}[ht]
\centering
% trim={<left> <lower> <right> <upper>}
\includegraphics[width=\linewidth,trim={0cm 8cm 2.5cm 0cm},clip]{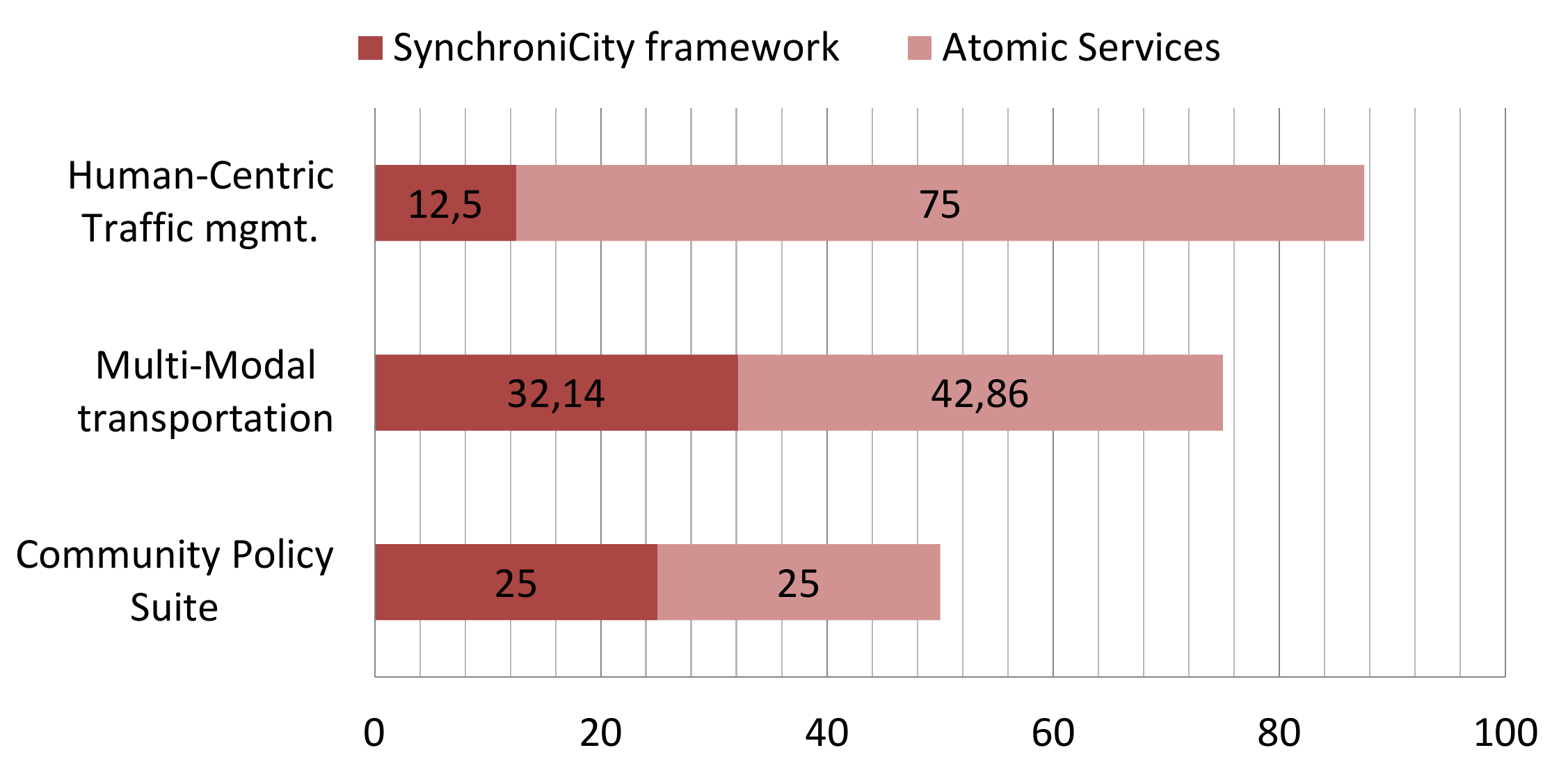}
\caption{Coverage in percentage of Function Requirements per application theme}
\label{fig:coverage}
\end{figure}

The roster of the atomic services, together with the features offered by the SynchroniCity framework, are already addressing a big percentage of application themes FRs of common interest among cities (see Section~\ref{sec:atomicservices}). Figure~\ref{fig:coverage} shows that the coverage goes from 50 to 87,5\%. More specifically: 7 FRs addressed out of 8 for the human-centric traffic management; 21 FRs addressed out of 28 for the multi-modal transportation; for the community policy suite only 4 out of 8 FRs are covered.

\begin{figure}[ht]
\centering
\includegraphics[width=\linewidth,trim={4cm 5.5cm 11cm 6cm},clip]{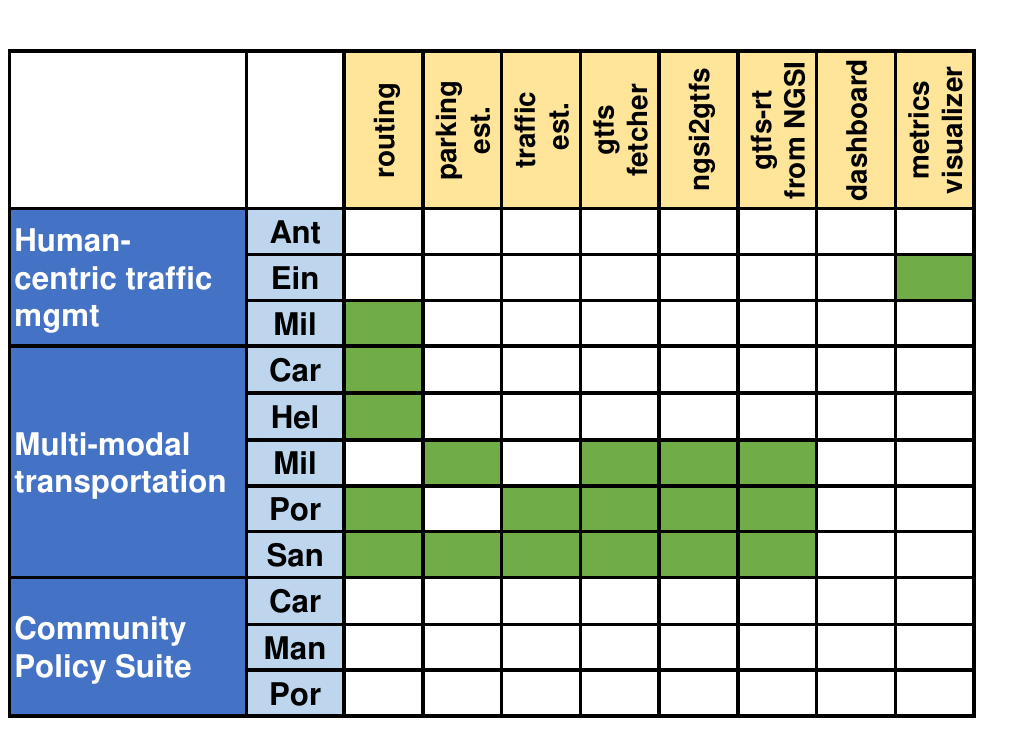}
\caption{Adoption of the atomic services per city services}
\label{fig:adoption}
\end{figure}

The atomic services have been offered as software components to smart city developers that would need to deploy and integrate them with their city services. Figure~\ref{fig:adoption} summarizes their adoption in each of the pilot cities involved in SynchroniCity. The most used atomic services are those that offer the most unique and complex features such as the data analytics estimators and the routing component together with integration atomic services. The offered visualization tools have been scarcely utilized. This is due to inclination of the city services developers to implement the graphical interface on their own in order to have a clear branding imprint with "look and feel" of the application. 
This explains also the lack of usage of atomic services in community policy suite applications since the visualization features are the most meaningful part.
Nevertheless, even if not displayed in Figure~\ref{fig:adoption}, the dashboard and the metrics visualizer have been utilized for fast prototyping and IoT data and devices discovering at city services development phase. 

\begin{figure}[ht]
\centering
\includegraphics[width=\linewidth,trim={0cm 13.5cm 4cm 0cm},clip]{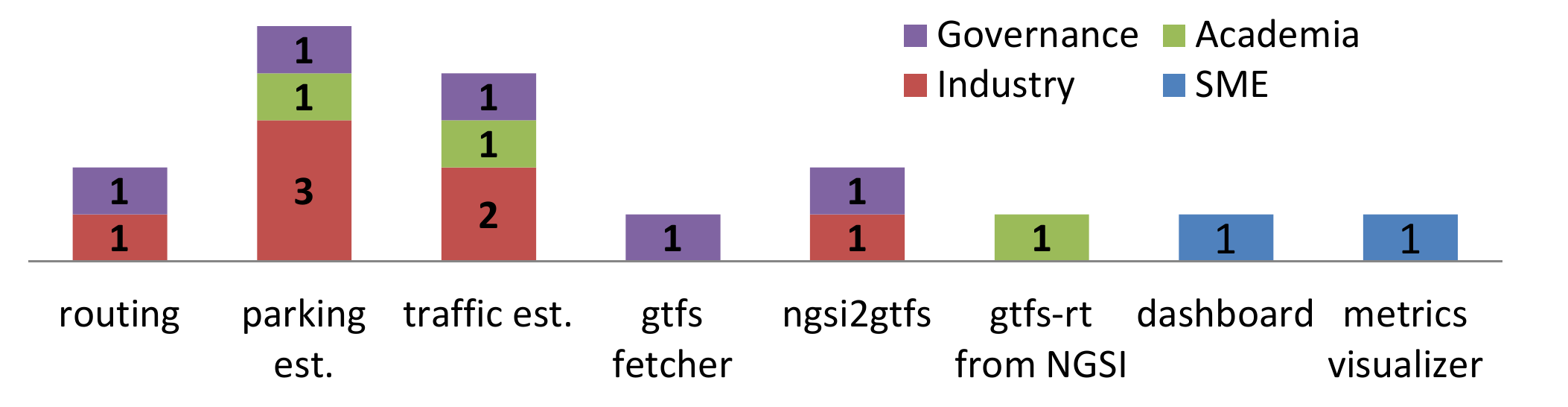}
\caption{Distribution of atomic service contributions per stakeholders}
\label{fig:contribution}
\end{figure}

Finally figure~\ref{fig:contribution} shows the distribution of contribution and cooperation amongst governance institutes, academia, industry and small and medium enterprises (SMEs).

%%%%%%%%%%%%%%%%%%%%%%%%%%%%%%%%%%%%%%%%%
%%%%% Conclusions %%%%%%%%%%%%%%%%%%%%%%
%%%%%%%%%%%%%%%%%%%%%%%%%%%%%%%%%%%%%%%%%
\section{Conclusions}

In this article we have illustrated the collaboration between 8 pilot cities to share efforts for smart city services and build up a sustainable ecosystem. In particular, we have depicted the collaboration methodology used and its application, based on the core concept of atomic service that is a single functional component to be re-used in more than one IoT application. A questionnaire has been implemented and issued to cities and technical partners to gather information regarding previous experiences and actual city needs. The analysis of the answers brought to the identification of required atomic functional blocks, namely atomic services, to be re-used by multiple stakeholders. Further atomic services have been identified by comparing designs of the planned city services.

We then introduce the design and architectures of the chosen atomic services, and of a city service made of a combination of multiple services. Finally, we evaluate the outcome of our approach depicting the actual adoption of the atomic services. Besides, we characterize the collaboration amongst different typology of parties (e.g. academia, industry, SMEs).

The finalization of the smart city services within SynchroniCity is still ongoing and it is foreseen the identification of new atomic services (3 are already on the process to be published as such). In order to keep alive these discussions we have activated a community tool\footnote{https://gitlab.com/synchronicity-atomicservicesevolution}. Finally, new pilot cities and city services developers are already joining the SynchroniCity project, pushing for further evolution of the ecosystem.

\section*{Acknowledgment}
This work has been partially funded by the European Union's Horizon 2020 Programme under Grant Agreement No. 732240 SynchroniCity (Delivering an IoT enabled Digital Single Market for Europe and Beyond). The content of this paper does not reflect the official opinion of the European Union. Responsibility for the information and views expressed therein lies entirely with the authors.

\bibliographystyle{IEEEtran}
% Generated by IEEEtran.bst, version: 1.14 (2015/08/26)

\end{document}